\documentclass[%
 reprint,
 amsmath,amssymb,
 aps,
]{revtex4-2}

\usepackage{hyperref}
\usepackage[utf8]{inputenc}
\usepackage{newunicodechar}
\newunicodechar{ }{\,}

\usepackage{graphicx}
\usepackage{dcolumn}
\usepackage{amsmath}
\usepackage{supertabular}
\usepackage{multirow}
\usepackage{indentfirst} 

\usepackage{lineno}

\usepackage{color}
\usepackage{cancel}
\usepackage{ulem}
\usepackage{array}
\usepackage{tabularx}

\usepackage{amssymb}
\usepackage{xcolor}

\usepackage{setspace}
\makeatletter
\def\section{\@startsection{section}{1}{\z@}%
  {-3.5ex \@plus -1ex \@minus -.2ex}%
  {2.3ex \@plus.2ex}%
  {\normalfont\large\bfseries\centering}}
\makeatother

\begin{document}

\title{Effects of femtoscopic correlations on spin-spin correlation measurements}

\bigskip

\author{Bijun Fan}
\affiliation{Key Laboratory of Quark and Lepton Physics (MOE) and Institute of Particle Physics, 
Central China Normal University, Wuhan 430079, China}

\author{Like Liu}
\email[Corresponding author, ]{likeliu@sdu.edu.cn}
\affiliation{Institute of Frontier and Interdisciplinary Science,
Key Laboratory of Particle Physics and Particle Irradiation (MOE),
Shandong University, QingDao, Shandong 266237, China}

\author{Shusu Shi}
\email[Corresponding author, ]{shiss@ccnu.edu.cn}
\affiliation{Key Laboratory of Quark and Lepton Physics (MOE) and Institute of Particle Physics, 
Central China Normal University, Wuhan 430079, China}

\date{\today}
\begin{abstract}
Hyperon spin correlations serve as sensitive probes of spin dynamics in high-energy collisions, yet their extraction from weak-decay angular distributions can be contaminated by femtoscopic effects due to quantum statistics and final-state interactions.
In this work, we quantitatively assess this contamination for $\Lambda\Lambda$ and $\Lambda\bar{\Lambda}$ pairs using the AMPT model combined with spin-dependent weights from the Lednick\'y--Lyuboshits formalism.
Because the singlet and triplet spin configurations contribute differently to the decay-angle distribution, femtoscopic weighting induces an apparent angular modulation, creating a fake correlation signal even when no intrinsic spin correlation is present.
We find that the induced bias can become substantial in the low-$q_{\mathrm{inv}}$ region, where the combined femtoscopic effect reaches a magnitude comparable to that of the preliminary CMS measurements.
Our results establish a framework for evaluating such systematics, highlighting that femtoscopic corrections must be carefully considered in future differential analyses that emphasize the low-relative-momentum region.
\end{abstract}

\maketitle

\section{Introduction}\label{sec.I}

Spin-spin correlation measurements provide a new way to probe spin dynamics in particle production and have recently attracted increasing interest in high-energy physics. 
The observation of quantum entanglement in top-quark pairs~\cite{ATLAS:2024TopEntanglement}, together with recent measurements of spin correlations in hyperon pairs~\cite{STAR:2025njp, CMS:Preliminary}, has opened new opportunities to study spin phenomena in particle production from both quantum-information and hadronization perspectives.
In the strange-baryon sector, the BESIII Collaboration has tested local realism using entangled $\Lambda\bar{\Lambda}$ pairs produced in $e^{+}e^{-}$ collisions~\cite{BESIII:2025LocalRealismLambda}, while STAR and CMS have reported spin-spin correlations for $\Lambda\bar{\Lambda}$, $\Lambda\Lambda$, and $\bar{\Lambda}\bar{\Lambda}$ pairs in high-energy $p$+$p$ collisions~\cite{STAR:2025njp, CMS:Preliminary}. 
These measurements provide new opportunities to study the spin structure of strangeness production, for which the $\Lambda$ hyperon plays a particularly useful role because its spin is sensitive to the polarization of its constituent strange quark~\cite{Ellis:2011va, Burkardt:1993zh, Ma:2001rm}.

The $\Lambda$ hyperon is particularly well suited for spin-correlation studies because its parity-violating weak decay is self-analyzing~\cite{Lee:1957,OPAL:1998, HERMES:2007,STAR:2017GlobalPolarization,BESIII:2019PolarizationEntanglement,HyperCP:2004}. The momentum direction of the daughter proton (antiproton) measured in the parent hyperon rest frame is correlated with the parent spin through the weak-decay asymmetry parameter $\alpha$. 
The spin correlation is therefore inferred statistically from the angular distribution of the decay products rather than measured directly.
In particular, the opening angle $\theta^{*}$ between the daughter (anti)protons in their respective parent-hyperon rest frames provides the angular observable used to extract the spin-spin correlation.

Despite their long-standing role as a sensitive probe of particle-emission sources and low-energy two-particle interactions, femtoscopic correlations arising from quantum statistics and final-state interactions~\cite{HanburyBrown:1956,Goldhaber:1960,Koonin:1977,Pratt:1984,Lisa:2005dd,Lednicky:2003mq} have been largely overlooked in the context of spin-spin correlation measurements.
For identical hyperons, the Pauli exclusion principle leads to a momentum-dependent relative weighting of the spin-singlet and spin-triplet channels at small relative momentum, as demonstrated experimentally through the momentum dependence of the spin fractions measured by the ALEPH Collaboration~\cite{ALEPH:1999ibh}. More generally, final-state interactions can also depend on the spin channel and modify the correlation functions of both identical and non-identical hyperon pairs. Since the spin information is statistically inferred from the decay-angle distributions of the daughter (anti)protons, such spin-dependent femtoscopic effects can change the relative weighting of hyperon pairs with different spin configurations. This, in turn, can modify the measured $\cos\theta^{*}$ distribution and bias the experimentally extracted spin-correlation observable if not properly taken into account. Quantifying such effects is therefore important for reliable spin-correlation measurements.

In this paper, we investigate the impact of femtoscopic effects on $\Lambda$-hyperon spin-spin correlation measurements using AMPT-generated events with femtoscopic correlation weights calculated from the analytical Lednick\'y--Lyuboshits formalism~\cite{Lednicky:1981su}.
This framework allows us to study Pauli suppression in $\Lambda\Lambda$ pairs and strong-interaction effects in both identical and non-identical hyperon pairs, including $\Lambda\bar{\Lambda}$.
The spin-state weights for each hyperon pair are determined from the decay angular distributions following the Alexander--Lipkin formalism~\cite{Alexander:1995nd} and combined with spin-dependent femtoscopic correlation functions to quantify the resulting modification of the extracted spin-correlation signal.
The analysis is performed for $p$+$p$ collisions at $\sqrt{s}=13$~TeV and $p$+Pb collisions at $\sqrt{s_{\mathrm{NN}}}=8.16$~TeV, and the model results are compared with the available preliminary CMS measurements.

The paper is organized as follows. 
Section II describes the construction of spin-state weights and the implementation of the femtoscopic correlation weighting. 
Section III quantifies the modification of the $\cos\theta^{*}$ distributions and the resulting bias in the extracted spin-correlation observable and compares the model results with preliminary CMS measurements and discusses the implications. Section IV summarizes our findings.

\section{Method}\label{sec.II}

In this work, the influence of the HBT effect on hyperon spin-correlation measurements is investigated within the framework of the AMPT model. The analysis consists of four main steps. First, $\Lambda(\bar{\Lambda})$ hyperons are generated using the AMPT event generator. Their weak decays are then simulated to construct the spin-correlation observable based on the decay (anti)protons angular distributions. Subsequently, for each hyperon pair, the spin-dependent femtoscopic correlation function calculated at the corresponding relative momentum $q_{\mathrm{inv}}$ is assigned as a correlation-function (CF) weight using the analytical Lednický--Lyuboshits formalism. Finally, the weighted angular distributions are analyzed to quantify the influence of the HBT effect on the extracted spin-correlation signal. The details of each step are described in the following subsections.

\subsection{AMPT event generator}

The event samples used in this work are generated with the string-melting A Multi-Phase Transport (AMPT) model~\cite{Lin:2004en}, using an improved version of the model~\cite{Zhang:2019utb, Zhang:2022fum,Zhang:2024zga}. The model version and parameter settings are the same as those used in our previous work~\cite{Fan:2026ife}.
Two collision systems are studied, namely $p$+$p$ collisions at $\sqrt{s}=13$ TeV and $p$+Pb collisions at $\sqrt{s_{\mathrm{NN}}}=8.16$ TeV.

The AMPT model describes relativistic heavy-ion and hadronic collisions through four successive stages: initial particle production from HIJING~\cite{Wang:1991hta}, partonic scatterings implemented by Zhang's Parton Cascade (ZPC)~\cite{Zhang:1997ej}, hadronization via the quark coalescence mechanism in the string-melting scenario, and subsequent hadronic rescatterings described by A Relativistic Transport (ART) model~\cite{Li:1995pra}. The generated final-state $\Lambda$ and $\bar{\Lambda}$ hyperons are used as the input particles for the present analysis.

In the AMPT event record, $\Lambda$ and $\bar{\Lambda}$ hyperons remain stable and therefore do not undergo weak decays during the event generation. Since the spin-correlation observable is defined through the momentum directions of the decay proton (antiproton), the weak decay of each $\Lambda(\bar{\Lambda})$ is performed during the analysis stage using the \texttt{TPythia6Decayer} package implemented in ROOT, based on PYTHIA~\cite{Sjostrand:1993yb}. The decay channels $\Lambda \rightarrow p+\pi^{-}$ and $\qquad
\bar{\Lambda} \rightarrow \bar{p}+\pi^{+}$ are simulated according to the corresponding decay table. Among all decay products, only the final-state proton (antiproton) is retained. To match the experimental acceptance, the $\Lambda(\bar{\Lambda})$ is required to satisfy $0.8<p_{\mathrm{T}}<6.0~\mathrm{GeV}/c$ and $|\eta|<2.4$.
The momentum direction of the accepted daughter proton (antiproton), evaluated in the rest frame of its parent hyperon, is subsequently used to construct the decay angle $\theta^{*}$ introduced in the following subsection.

\subsection{Construction of spin-state weights}

The spin state of an individual hyperon pair cannot be directly identified experimentally. Instead, the relative contributions of the spin-singlet ($S=0$) and spin-triplet ($S=1$) configurations are inferred statistically from the angular correlations of the decay products. As described in the previous subsection, $\theta^{*}$ is defined as the opening angle between the momentum directions of the daughter proton (antiproton) from the weak decays of the two parent hyperons, with each momentum evaluated in the rest frame of its corresponding parent hyperon.

The method follows the formalism proposed by Alexander and Lipkin~\cite{Alexander:1995nd}, in which the parity-violating weak decay of the $\Lambda$ hyperon is exploited as a spin analyzer. For a polarized hyperon, the angular distribution of the daughter proton (antiproton) is given by
\begin{equation}
\frac{dN}{d\Omega}
=
\frac{1}{4\pi}
\left(
1+\alpha\,\mathbf{P}\cdot\hat{\mathbf{n}}
\right),
\end{equation}
where $\mathbf{P}$ is the polarization vector of the parent hyperon, $\hat{\mathbf{n}}$ is the unit vector along the momentum direction of the daughter proton (antiproton), and $\alpha$ is the weak-decay asymmetry parameter.

For a pair of spin-$1/2$ hyperons, the spin correlation can be characterized by the two-particle spin-correlation tensor,
\begin{equation}
T_{ik}
=
\left\langle
\sigma_{i}^{(1)}
\sigma_{k}^{(2)}
\right\rangle,
\end{equation}
where $\sigma_{i}^{(1)}$ and $\sigma_{k}^{(2)}$ are the components of the Pauli spin operators acting on the two hyperons. Following the spin-correlation formalism of Ref.~\cite{Lednicky:2001yw}, we define the normalized spin-correlation observable as
\begin{equation}
P_{\Lambda\Lambda}
\equiv
\frac{1}{3}\operatorname{Tr}(T),
\end{equation}
where $\operatorname{Tr}(T)=T_{11}+T_{22}+T_{33}$. For the spin-singlet and spin-triplet sectors, the trace of the spin-correlation tensor takes the values $-3$ and $+1$, respectively, giving
\begin{equation}
P_{\Lambda\Lambda}
=
\begin{cases}
-1, & S=0,\\[4pt]
\frac{1}{3}, & S=1.
\end{cases}
\end{equation}
Thus, the singlet and triplet sectors correspond to distinct values of the spin-correlation observable.

The parity-violating weak decay allows the relative contributions of these two spin sectors to be inferred from the opening-angle distribution. Following the Alexander--Lipkin formalism~\cite{Alexander:1995nd}, the angular distributions for pure singlet and triplet configurations are
\begin{equation}
\frac{dN_{(S=0)}}{d\cos\theta^{*}}
\propto
1-\alpha_{1}\alpha_{2}\cos\theta^{*},
\end{equation}
and
\begin{equation}
\frac{dN_{(S=1)}}{d\cos\theta^{*}}
\propto
1+\frac{1}{3}\alpha_{1}\alpha_{2}\cos\theta^{*}.
\end{equation}

The singlet sector contains one spin state, whereas the triplet sector contains three. Their statistical multiplicities therefore give the factors $1/4$ and $3/4$ for an unpolarized pair. For a given hyperon pair, the corresponding probabilities for the singlet and triplet sectors are obtained from the relative contributions of the two angular distributions,
\begin{equation}
W_{0}
=
\frac{
\frac{1}{4}\frac{dN_{(S=0)}}{d\cos\theta^{*}}
}{
\frac{1}{4}\frac{dN_{(S=0)}}{d\cos\theta^{*}}
+
\frac{3}{4}\frac{dN_{(S=1)}}{d\cos\theta^{*}}
},
\qquad
W_{1}=1-W_{0}.
\end{equation}

\begin{figure}[!htbp]
  \centering   
  \includegraphics[width=0.45\textwidth]{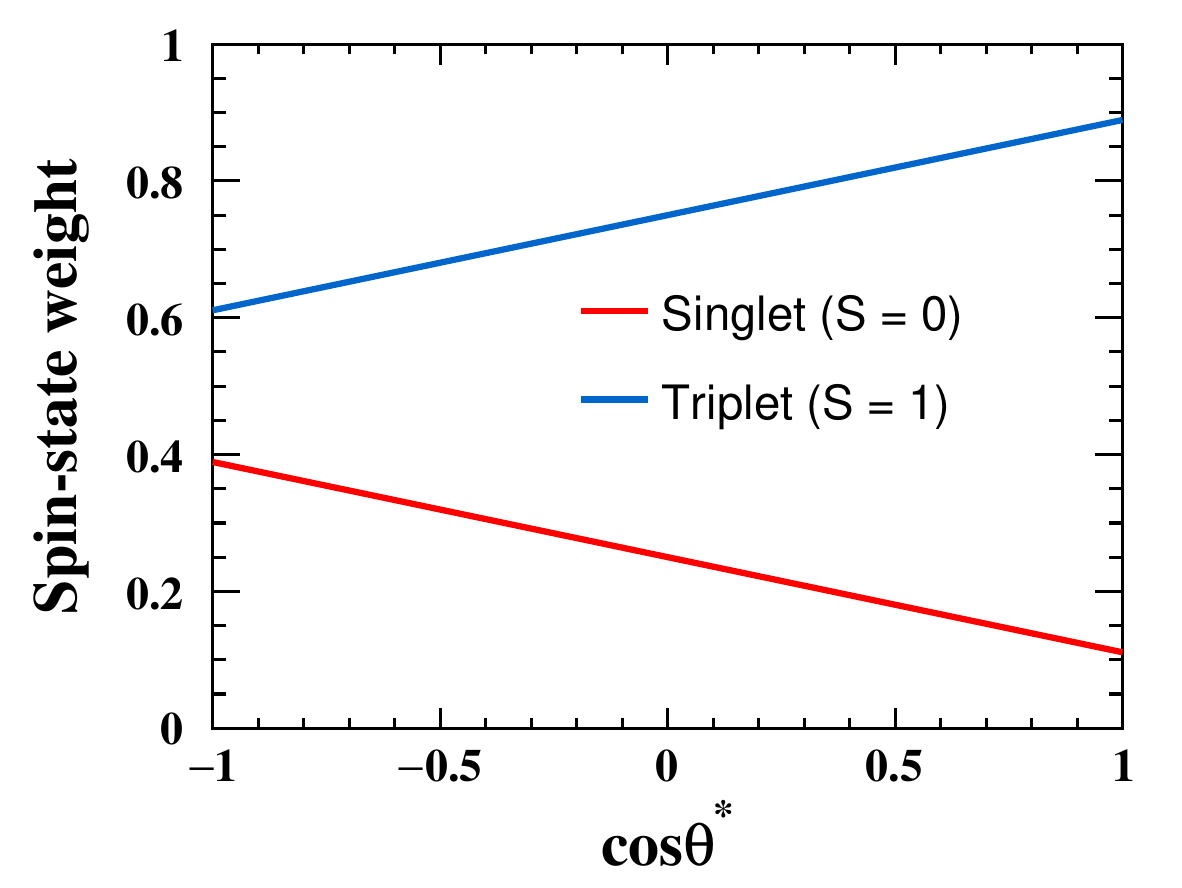}
   \caption{Spin-state weights for the spin-singlet ($S=0$) and spin-triplet ($S=1$) configurations as functions of $\cos\theta^{*}$ for $\Lambda\Lambda$ pairs. 
   The weights for $\Lambda\bar{\Lambda}$ pairs are slightly different because of the different weak-decay parameters of the $\Lambda$ and $\bar{\Lambda}$ hyperons.
   }
    \label{spin_weight_vs_costheta}
\end{figure}

In the present work, the weak-decay asymmetry parameters $\alpha_{\Lambda}=0.746$ and $\alpha_{\bar{\Lambda}}=-0.758$ are adopted from the 2026 Particle Data Group~\cite{ParticleDataGroup:2026aaa}. For $\Lambda\Lambda$ pairs, the resulting spin-state weights are
\begin{equation}
W_{0}
=
0.25-0.139129\cos\theta^{*},
\end{equation}
and
\begin{equation}
W_{1}
=
0.75+0.139129\cos\theta^{*},
\end{equation}
These weights are used throughout the $\Lambda\Lambda$ analysis. The same formalism is applied to $\Lambda\bar{\Lambda}$ pairs using the corresponding decay asymmetry parameters, resulting in different numerical weights because $\alpha_{\Lambda}$ and $\alpha_{\bar{\Lambda}}$ have different values and signs. The resulting spin-state weights are shown in Fig.~\ref{spin_weight_vs_costheta}. These weights represent the statistical probabilities assigned to the singlet and triplet sectors for a given pair according to its $\cos\theta^{*}$ value and are subsequently combined with the spin-dependent femtoscopic correlation functions described below.

\subsection{Spin-dependent femtoscopic correlation-function weights}

To incorporate femtoscopic effects into the spin-correlation analysis, a correlation-function (CF) weight is assigned to each hyperon pair. 
For a given pair, the weight is constructed from the correlation functions of the spin-singlet ($S=0$) and spin-triplet ($S=1$) channels according to their corresponding spin-state probabilities,
\begin{equation}
C_{\mathrm{pair}}(q_{\mathrm{inv}})
=
W_{0}C_{S=0}(q_{\mathrm{inv}})
+
W_{1}C_{S=1}(q_{\mathrm{inv}}).
\label{eq:general_CF_weight}
\end{equation}
The channel-dependent correlation functions are calculated using the analytical Lednick\'y--Lyuboshits formalism~\cite{Lednicky:1981su,Lednicky:2001yq,Lednicky:2003mq}.

For identical $\Lambda\Lambda$ pairs, both quantum statistics (QS) and the strong interaction (SI) contribute to the femtoscopic correlation. 
For a Gaussian source with radius $r_{0}$, the QS correlation functions for the singlet and triplet channels are
\begin{equation}
C^{\mathrm{QS}}_{S=0}(q_{\mathrm{inv}})
=
1+\exp(-r_{0}^{2}q_{\mathrm{inv}}^{2}),
\label{eq:QS_singlet}
\end{equation}
\begin{equation}
C^{\mathrm{QS}}_{S=1}(q_{\mathrm{inv}})
=
1-\exp(-r_{0}^{2}q_{\mathrm{inv}}^{2}).
\label{eq:QS_triplet}
\end{equation}
The SI contribution is calculated using the standard Lednick\'y--Lyuboshits formalism~\cite{Lednicky:1981su,Lednicky:2003mq} with the scattering length $f_{0}$ and effective range $d_{0}$ as input parameters. 
For the spin-singlet channel, the corresponding correlation function is
\begin{align}
C_{S=0}^{\mathrm{SI}}(k^{*})
=&\,
1+
\frac{1}{2}
\frac{|f(k^{*})|^{2}}{r_{0}^{2}}
\left(
1-\frac{d_{0}}{2\sqrt{\pi}r_{0}}
\right)
\nonumber\\
&+
\frac{2\operatorname{Re}[f(k^{*})]}
{\sqrt{\pi}r_{0}}
F_{1}(2k^{*}r_{0})
-
\frac{\operatorname{Im}[f(k^{*})]}
{r_{0}}
F_{2}(2k^{*}r_{0}),
\label{eq:SI_CF}
\end{align}
where $k^{*}=q_{\mathrm{inv}}/2$ and $f(k^{*})$, $F_{1}$, and $F_{2}$ follow the standard definitions of the Lednick\'y--Lyuboshits formalism~\cite{Lednicky:1981su,Lednicky:2003mq}.
For the present $\Lambda\Lambda$ calculation, the strong interaction is included only in the spin-singlet channel due to the Pauli exclusion principle, while the triplet channel is affected only by QS.

Three femtoscopic weighting schemes are considered for $\Lambda\Lambda$ pairs. 
In the QS-only case, the pair weight is constructed from the QS correlation functions in Eq.~\eqref{eq:QS_singlet} and Eq.~\eqref{eq:QS_triplet}. 
In the SI-only case, the spin-singlet channel is weighted by $C_{S=0}^{\mathrm{SI}}$, while the triplet channel has unit weight,
\begin{equation}
C_{\mathrm{pair}}^{\mathrm{SI}}
=
W_{0}C_{S=0}^{\mathrm{SI}}
+
W_{1}.
\label{eq:pair_CF_SI}
\end{equation}
For the SI+QS case, the combined quantum-statistical and strong-interaction correlation functions are calculated following the Lednick\'y--Lyuboshits formalism~\cite{Lednicky:1981su,Lednicky:2003mq}, with the strong interaction included in the singlet channel and QS included in both spin channels.

\begin{table}[tb]
\centering
\caption{Source radii and spin-dependent interaction parameters used in the analytical Lednick\'y--Lyuboshits calculations~\cite{STAR:2014dcy,ALICE:2018ysd,ALICE:2019eol,ALICE:2019igo}.
All parameters are given in units of fm.}
\label{tab:LL_parameters}
\begin{tabular}{llccccc}
\hline\hline
Collision & Pair & Spin & $r_{0}$ & $\mathrm{Re}(f_{0})$ &
$\mathrm{Im}(f_{0})$ & $d_{0}$ \\
\hline

\multirow{3}{*}{$p$+$p$} & $\Lambda\Lambda$ & $S=0$ &
\multirow{3}{*}{1.5} & 1.0 & -- & 5.0 \\
& $\Lambda\bar{\Lambda}$ & $S=0$ & & $-2.0$ & 1.20 & 5.0 \\

& $\Lambda\bar{\Lambda}$ & $S=1$ & & $-0.3$ & 0.30 & 2.0 \\

\hline

\multirow{3}{*}{$p$+Pb} & $\Lambda\Lambda$ & $S=0$ &
\multirow{3}{*}{2.0} & 1.0 & -- & 5.0 \\

& $\Lambda\bar{\Lambda}$ & $S=0$ & & $-2.0$ & 1.20 & 5.0 \\

& $\Lambda\bar{\Lambda}$ & $S=1$ & & $-0.3$ & 0.30 & 2.0 \\

\hline\hline
\end{tabular}
\end{table}

For non-identical $\Lambda\bar{\Lambda}$ pairs, quantum statistics is absent, while the strong interaction contributes to both spin channels. 
The scattering length is therefore allowed to be complex,
\begin{equation}
f_{0}
=
\mathrm{Re}(f_{0})
+
i\,\mathrm{Im}(f_{0}),
\end{equation}
to account for annihilation and other inelastic coupled-channel processes~\cite{Schneider:1992}.
The pair weight is then
\begin{equation}
C_{\mathrm{pair}}^{\mathrm{SI}}
=
W_{0}C_{S=0}^{\mathrm{SI}}
+
W_{1}C_{S=1}^{\mathrm{SI}}.
\label{eq:pair_CF_LamALam}
\end{equation}
Since quantum statistics is absent for $\Lambda\bar{\Lambda}$ pairs, the SI-only and SI+QS weighting schemes are identical for this system.

The source radius and interaction parameters used in the Lednick\'y--Lyuboshits calculations are summarized in Table~\ref{tab:LL_parameters}. 
The adopted interaction parameters are motivated by previous experimental femtoscopic measurements~\cite{STAR:2014dcy,ALICE:2018ysd,ALICE:2019eol,ALICE:2019igo}, which constrain the corresponding hyperon--hyperon interactions, although the available measurements do not precisely determine the interaction parameters separately for the spin-singlet and spin-triplet channels.
In addition to the femtoscopically weighted samples, a reference sample without any CF weight (No-weight) is also analyzed.
Thus, four weighting scenarios are considered throughout this work: No-weight, QS-only, SI-only, and SI+QS.

\begin{figure*}[!htbp]
  \centering   
  \includegraphics[width=0.75\textwidth]{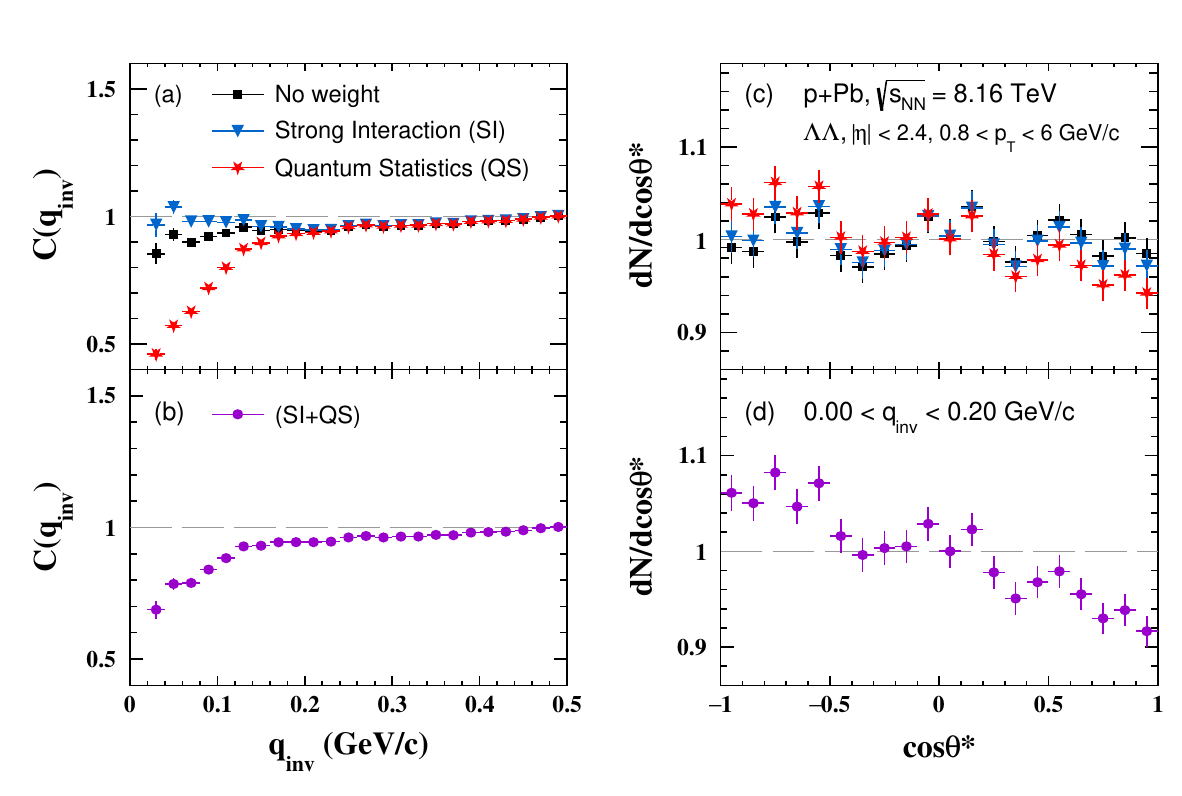}
   \caption{Distributions of the correlation function $C(q_{\mathrm{inv}})$ (left) and $dN/d\cos\theta^{*}$ (right) for $\Lambda\Lambda$ pairs in $p+Pb$ collisions at $\sqrt{s_{\mathrm{NN}}}=8.16$ TeV. The $dN/d\cos\theta^{*}$ distributions are constructed for pairs with $0<q_{\mathrm{inv}}<0.20$ GeV/$c$. The upper panels (a) and (c) compare three cases: without CF weight (No weight), including only the strong interaction (SI), and including only the quantum statistics (QS). The lower panels (b) and (d) show the results obtained by including both effects simultaneously.
   }
    \label{costheta_CF_LamLam_q0to200MeV}
\end{figure*}

\subsection{Extraction of the spin-correlation signal}

For each hyperon pair, the weight described in the previous subsection is applied before filling the $dN/d\cos\theta^{*}$ distribution. Unless otherwise specified, the angular distributions used to study the HBT effect are constructed from pairs with $0<q_{\mathrm{inv}}<0.2$ GeV/$c$, where the femtoscopic effect is most pronounced.

To investigate how the HBT effect modifies the experimentally extracted spin-correlation signal, an artificial spin correlation is introduced into the generated event sample. Instead of modifying the underlying event generation, the injected signal is implemented statistically through an acceptance--rejection procedure based on the decay-angle distribution. For a given input polarization strength $P_{\Lambda\Lambda}^{\mathrm{in}}$, each hyperon pair is accepted with a probability
\begin{equation}
P_{\mathrm{acc}}
=
\frac{1+\alpha_{1}\alpha_{2} P_{\Lambda\Lambda}^{\mathrm{in}}\cos\theta^{*}}
{1+\left|\alpha_{1}\alpha_{2} P_{\Lambda\Lambda}^{\mathrm{in}}\right|},
\end{equation}
where $\alpha_{1}\alpha_{2}$ is a proportionality constant relating the injected polarization strength to the slope of the $\cos\theta^{*}$ distribution. The denominator guarantees that the acceptance probability remains within the physical interval $0\le P_{\mathrm{acc}}\le1$ for all generated pairs. The same underlying event sample is used for all values of $P_{\Lambda\Lambda}^{\mathrm{in}}$. For each chosen input value, the acceptance--rejection procedure is applied independently using the corresponding acceptance probability, thereby producing samples with different injected spin-correlation strengths without regenerating the underlying events.

The weighted $dN/d\cos\theta^{*}$ distributions are subsequently fitted with a linear function,
\begin{equation}
\frac{dN}{d\cos\theta^{*}}
=
N_{0}
\left(
1+\alpha_{1}\alpha_{2} P_{\Lambda\Lambda}^{\mathrm{out}}\cos\theta^{*}
\right),
\end{equation}
from which the extracted spin-correlation strength, $P_{\Lambda\Lambda}^{\mathrm{out}}$, is determined. By comparing the extracted value with the injected polarization strength, the influence of different femtoscopic weighting schemes on the measured spin-correlation signal can be quantified.

\section{Results and Discussion}\label{sec.III}

Figure~\ref{costheta_CF_LamLam_q0to200MeV} presents the correlation functions together with the corresponding $\cos\theta^{*}$ distributions in the range $0<q_{\mathrm{inv}}<0.20$ GeV/$c$ for $\Lambda\Lambda$ pairs under four different scenarios in $p+Pb$ collisions at $\sqrt{s_{\mathrm{NN}}}=8.16$ TeV: without any correlation weight, including only the strong interaction (SI), including only quantum statistics (QS), and including both effects (SI + QS). The CF weights are evaluated using the analytical Lednický--Lyuboshits formalism mentioned before and are assigned to each pair according to its relative momentum $q_{\mathrm{inv}}$.

Without any weight, the $\cos\theta^{*}$ distribution is nearly flat, indicating the absence of an intrinsic spin-correlation signal in the generated events. The corresponding unweighted correlation function is slightly below unity at low $q_{\rm inv}$ and gradually approaches unity at larger $q_{\rm inv}$. This small residual deviation reflects non-femtoscopic correlations in the generated events, which can arise in small collision systems in addition to the femtoscopic correlations of interest~\cite{ALICE:2019eol}. The $dN/d\cos\theta^{*}$ distributions are constructed following the procedure used in the experimental measurement~\cite{CMS:Preliminary}, with the same kinematic selections applied to the reconstructed $\Lambda$ candidates.

After applying the CF weights, distinct modifications are observed. The strong interaction (SI) alone produces a relatively small modification of the correlation function and only a weak dependence on $\cos\theta^{*}$. In contrast, the inclusion of quantum statistics (QS) leads to a pronounced angular modulation, with an enhancement at negative $\cos\theta^{*}$ and a suppression at positive $\cos\theta^{*}$. When both effects are included, the resulting $\cos\theta^{*}$ distribution exhibits a behavior very similar to that obtained with QS alone, indicating that the observed angular modification is dominated by the quantum-statistical effect for $\Lambda\Lambda$ pairs. Thus, although the different weighting schemes modify the correlation function to different degrees, the angular dependence induced by SI is comparatively weak, whereas the inclusion of QS results in a pronounced modification of the $\cos\theta^{*}$ distribution.

\begin{figure*}[!htbp]
  \centering   
  \includegraphics[width=0.95\textwidth]{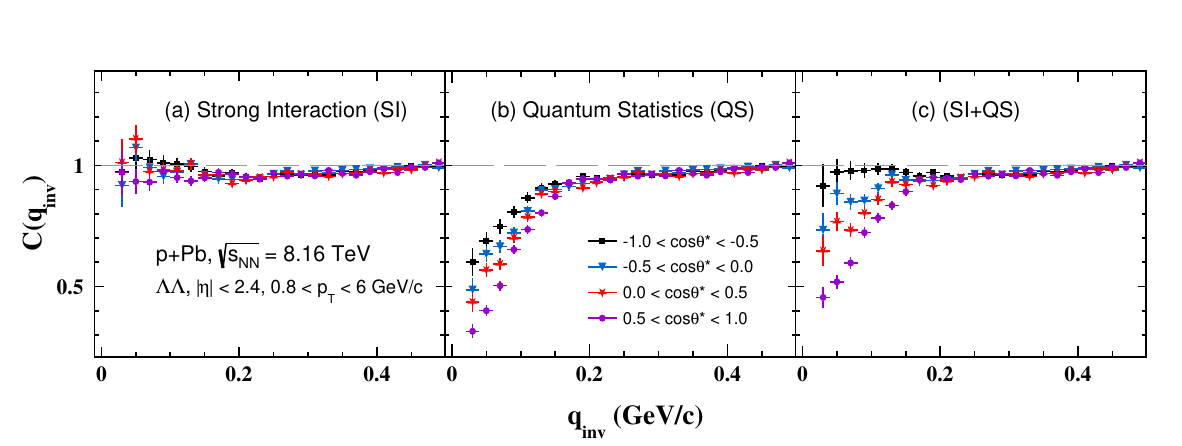}
   \caption{Correlation functions $C(q_{\mathrm{inv}})$ for $\Lambda\Lambda$ pairs in $p+Pb$ collisions at $\sqrt{s_{\mathrm{NN}}}=8.16$ TeV, separately constructed for four $\cos\theta^{*}$ intervals: $-1.0<\cos\theta^{*}<-0.5$, $-0.5<\cos\theta^{*}<0.0$, $0.0<\cos\theta^{*}<0.5$, and $0.5<\cos\theta^{*}<1.0$. The panels correspond to the cases including (a) only the strong interaction (SI), (b) only the quantum statistics (QS), and (c) both effects simultaneously, respectively.}
   \label{Cqinv_ctBin_LamLam_q0to200MeV}
\end{figure*}

To further investigate the origin of the angular dependence observed in Fig.~\ref{costheta_CF_LamLam_q0to200MeV}, the correlation functions are evaluated separately in four $\cos\theta^{*}$ intervals, as shown in Fig.~\ref{Cqinv_ctBin_LamLam_q0to200MeV}. This differential representation allows the dependence of the correlation function on the decay angle to be examined for the strong interaction, quantum statistics, and their combined effects.

For the SI-only case, the correlation functions in the four $\cos\theta^{*}$ intervals show only modest differences, indicating a relatively weak angular dependence. The weak angular dependence induced by the SI therefore leads to only a small modification of the $\cos\theta^{*}$ distribution after the weighting procedure, as observed in Fig.~\ref{costheta_CF_LamLam_q0to200MeV}. This explains why the SI-only case produces only a weak slope in the $dN/d\cos\theta^{*}$ distribution, even though the SI modifies the correlation function at low $q_{\rm inv}$.

A distinctly different behavior is observed when only the quantum statistical (QS) effect is included. The correlation functions exhibit a  dependence on $\cos\theta^{*}$, with the correlation strength increasing systematically from the most negative $\cos\theta^{*}$ interval to the most positive one.
Since the spin-state weights are functions of $\cos\theta^{*}$, different angular intervals correspond to different relative contributions from the spin-singlet and spin-triplet configurations. Consequently, when the correlation strengths associated with the two spin channels are different, pairs in different $\cos\theta^{*}$ intervals acquire different relative CF modifications. This spin-dependent femtoscopic weighting is therefore translated into an apparent modulation of the $\cos\theta^{*}$ distribution, giving rise to a fake spin-correlation signal even in the absence of an intrinsic spin correlation in the generated events.

The combined SI+QS case exhibits a behavior very similar to that of the QS-only case, indicating that the angular dependence of the correlation function is dominated by the quantum-statistical effect. More generally, these results demonstrate that a fake spin-correlation signal can arise when the correlation functions associated with different spin channels are different. Because the spin-state weights vary with $\cos\theta^{*}$, this spin dependence of the correlation function can be translated into an apparent angular dependence of the weighted $\cos\theta^{*}$ distribution. This provides the physical basis for the HBT-induced bias in the extracted spin-correlation signal and motivates the quantitative studies presented in the following section.


Having established the origin of the HBT-induced apparent spin-correlation signal, we next investigate how the HBT effect influences the extraction of an injected spin correlation. 
Because the AMPT events contain no intrinsic spin correlation, samples with prescribed input spin-correlation strengths are constructed using an acceptance--rejection procedure.
Each hyperon pair is accepted or rejected with a probability determined by its $\cos\theta^{*}$ value and the input polarization, thereby producing the target angular distribution.
The extracted polarization parameter, $P_{\Lambda_{1}\Lambda_{2}}^{\mathrm{fit}}$, is then obtained by fitting the weighted $dN/d\cos\theta^{*}$ distribution in the region $0<q_{\mathrm{inv}}<0.20$ GeV/$c$, where the HBT-induced modification is expected to be most pronounced. 

\begin{figure*}[htbp]
    \centering
    \includegraphics[width=0.65\textwidth]{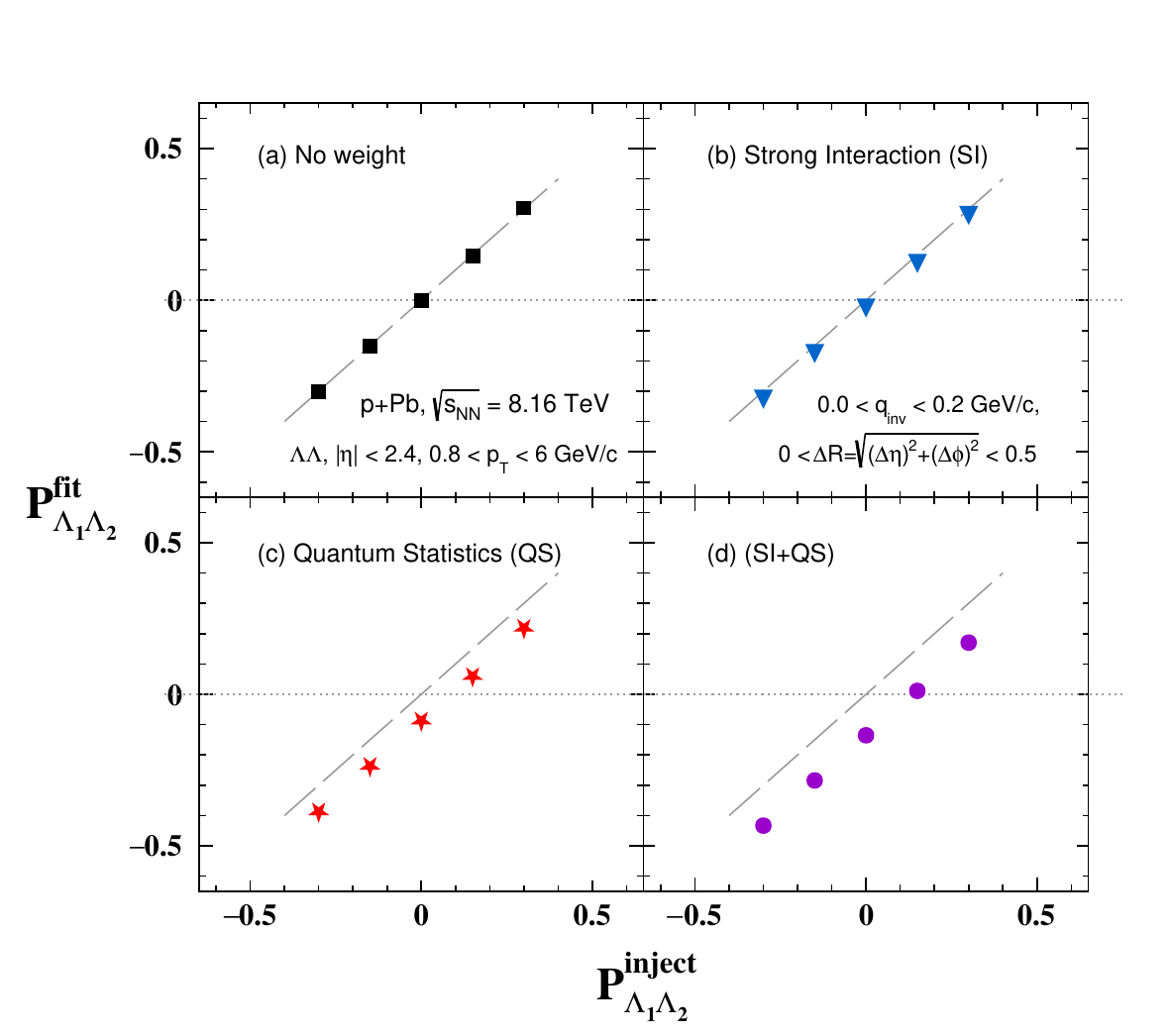}
    \caption{Comparison of the fitted and injected $\Lambda\Lambda$ spin-correlation strengths in $p+Pb$ collisions at $\sqrt{s_{\mathrm{NN}}}=8.16$ TeV for pairs with $0<\Delta R<0.5$ and $0<q_{\mathrm{inv}}<0.20$ GeV/$c$. 
    The panels represent, respectively, (a) the cases without CF weight (No weight), (b) including only the strong interaction (SI), (c) only the quantum statistics (QS), and (d) both effects simultaneously. The dashed diagonal line corresponds to $P_{\mathrm{fit}}=P_{\mathrm{inject}}$, providing a direct reference for evaluating possible biases introduced by different femtoscopic effects in the extraction of the spin-correlation signal.}
    \label{Pfit_vs_Pinject_drBin0_LamLam_q0to200MeV}
\end{figure*}

Figure~\ref{Pfit_vs_Pinject_drBin0_LamLam_q0to200MeV} compares the extracted and injected polarization parameters for four different weighting cases in the interval $0<\Delta R=\sqrt{(\Delta\eta)^2+(\Delta\phi)^2}<0.5$. The dashed diagonal represents the ideal case in which the extracted polarization is identical to the injected value. Without any CF weight, all data points lie along the diagonal, demonstrating that the fitting procedure accurately reproduces the injected spin-correlation strength. When only the strong interaction is included, the extracted polarization remains very close to the ideal relation, indicating that the strong interaction alone introduces only a negligible bias in the extracted spin-correlation signal. In contrast, noticeable deviations from the diagonal are observed when the quantum-statistical effect is included. 

The observed deviation originates from the angular dependence of the CF weight discussed in the previous section. As demonstrated in Fig.~\ref{Cqinv_ctBin_LamLam_q0to200MeV}, different $\cos\theta^{*}$ intervals acquire different effective CF modifications when the correlation functions of the two spin channels are different. Consequently, the CF weighting introduces an additional angular modulation into the measured $dN/d\cos\theta^{*}$ distribution. Since the spin-correlation strength is extracted from the slope of this angular distribution, the additional angular modulation introduced by the HBT effect modifies the fitted slope and consequently biases the extracted polarization parameter. The resulting deviation is not a simple constant offset; rather, it depends on the interplay between the angular modulation associated with the injected spin-correlation signal and the additional angular dependence introduced by the HBT effect. Depending on their relative magnitudes and signs, the HBT effect may therefore either increase or decrease the extracted spin-correlation strength.

An important observation is that the extracted polarization remains approximately linearly dependent on the injected polarization for all weighting cases. Thus, the HBT effect does not destroy the linear response of the measurement, but rather modifies the mapping between the injected and extracted spin-correlation strengths. This indicates that the HBT effect should be regarded as a systematic bias in the extraction procedure rather than as a random distortion. Such a bias is expected to be  particularly relevant in the low-$q_{\mathrm{inv}}$ and small-$\Delta R$ regions, where femtoscopic effects are most pronounced, and should therefore be properly evaluated in future experimental measurements.

\begin{figure*}[htbp]
    \centering
    \includegraphics[width=0.85\textwidth]{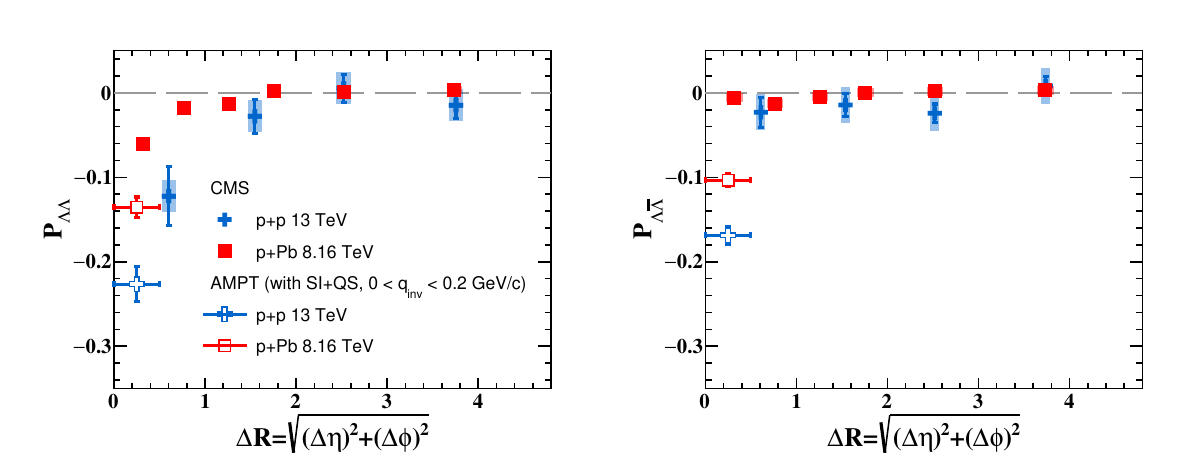}
    \caption{Comparison of the extracted spin-correlation strengths as a function of $\Delta R$ between the present model calculation and the CMS preliminary measurements~\cite{CMS:Preliminary}, for hyperons with $|\eta|<2.4$ and $0.8<p_{\rm T}<6$ GeV/c. The left and right panels show the results for $\Lambda\Lambda$ and $\Lambda\bar{\Lambda}$ pairs, respectively. The model calculations include the full final-state interaction (QS+SI), and the spin-correlation strengths are extracted from the $dN/d\cos\theta^{*}$ distributions using the pairs with $0<q_{\mathrm{inv}}<0.20$ GeV/$c$, in order to make the effect more obvious to observe. 
}
\label{Pfit_vs_deltaR_q0to2000MeV}
\end{figure*}

We next examine the HBT-induced modification of the extracted spin-correlation strength as a function of $\Delta R$ and compare the model results with the corresponding CMS preliminary measurements in $p$+$p$ and $p$+Pb collisions~\cite{CMS:Preliminary}, as shown in Fig.~\ref{Pfit_vs_deltaR_q0to2000MeV}. The model results are evaluated with $q_{\rm inv}<0.20$ GeV/$c$, consistent with the selection used in the preceding studies. Because $q_{\rm inv}$ and $\Delta R$ are strongly correlated in the generated samples, this low-$q_{\rm inv}$ selection restricts the model result to the smallest-$\Delta R$ interval.

For $\Lambda\Lambda$ pairs, the model predicts a negative spin-correlation strength in the smallest-$\Delta R$ interval, with a larger magnitude than the corresponding experimental values and therefore a more negative value. For $\Lambda\bar{\Lambda}$ pairs, the experimental results show a much weaker dependence on $\Delta R$, while the model also predicts a negative spin-correlation strength in the smallest-$\Delta R$ interval. These results indicate that the combined SI+QS femtoscopic weighting can produce a non-negligible modification of the extracted spin-correlation strength in the low-$q_{\rm inv}$ region. In the present calculation, this modification shifts the extracted spin-correlation strength toward more negative values for both $\Lambda\Lambda$ and $\Lambda\bar{\Lambda}$ pairs.

It is important to note, however, that the low-$q_{\rm inv}$ selection used in the model calculation above is not imposed in the current experimental analysis. When the full available $q_{\rm inv}$ range is used, the model gives a $\Lambda\Lambda$ spin-correlation strength close to zero, indicating that the net HBT-induced bias is small after integrating over the full relative-momentum range. Thus, the sizable modification observed in the low-$q_{\rm inv}$ calculation should not be interpreted as a direct estimate of the bias in the currently reported experimental result. Rather, it demonstrates that the spin-correlation observable can be sensitive to femtoscopic effects when the analysis emphasizes the low-relative-momentum region.

More generally, the direction and magnitude of the HBT-induced bias are not universal, but depend on the strength and direction of the $\cos\theta^{*}$ dependence of the correlation function. As discussed above, a stronger angular dependence of the CF leads to a larger modification of the extracted spin-correlation strength, while the sign of the modification is determined by the direction of this angular dependence relative to the injected spin-correlation signal. The negative shift observed for both $\Lambda\Lambda$ and $\Lambda\bar{\Lambda}$ pairs in the present calculation therefore reflects the specific angular dependence of the SI+QS correlation functions used here, rather than a universal tendency of femtoscopic effects to generate a negative spin-correlation signal.

The present study therefore provides a framework for evaluating femtoscopic contributions to hyperon spin-correlation measurements under different kinematic selections. For the full-$q_{\rm inv}$ range used in the current measurement, the model indicates that the overall HBT-induced bias is small. Nevertheless, the effect can become appreciable in the low-$q_{\rm inv}$ region, where femtoscopic correlations are strongest. This suggests that femtoscopic effects should be considered when evaluating systematic uncertainties in future spin-correlation measurements, particularly for analyses with explicit low-$q_{\rm inv}$ selections or other kinematic requirements that enhance the femtoscopic contribution.

\section{Summary}\label{sec.IV}

In this work, we investigated the impact of femtoscopic correlations on spin-correlation measurements of $\Lambda\Lambda$ and $\Lambda\bar{\Lambda}$ pairs. AMPT-generated events were combined with the Lednický--Lyuboshitz formalism to incorporate quantum-statistical and strong-interaction effects, with the corresponding correlation-function weights assigned to individual hyperon pairs. Because the spin-singlet and spin-triplet states have different femtoscopic correlation functions and contribute with $\cos\theta^{*}$-dependent weights to the weak-decay angular distribution, the femtoscopic weighting can introduce an additional angular modulation into $dN/d\cos\theta^{*}$. As a result, an apparent spin-correlation signal can be generated even when the underlying event sample contains no intrinsic spin correlation.

We further investigated the effect on the extraction of an injected spin-correlation signal. The extracted spin-correlation strength remains approximately linearly related to the injected value after femtoscopic weighting, while the mapping between the two is modified. The magnitude and direction of this modification are determined by the $\cos\theta^{*}$ dependence of the femtoscopic correlation function and are therefore not universal. 

For the $\Lambda\Lambda$ and $\Lambda\bar{\Lambda}$ systems considered here, the combined SI+QS weighting produces a negative shift of the extracted spin-correlation strength in the low-$q_{\rm inv}$ region, with the effect becoming particularly pronounced at small $\Delta R$. A comparison with the CMS preliminary measurements shows that, in this low-$q_{\rm inv}$ region, the predicted modification for $\Lambda\Lambda$ is of comparable magnitude to the experimentally observed negative spin-correlation strength, while a negative modification is also obtained for $\Lambda\bar{\Lambda}$.
When the $0<q_{\mathrm{inv}}<0.20$ GeV/$c$ selection is removed, the magnitude of the negative shift decreases because low-$q_{\rm inv}$ pairs constitute a small fraction of the entire $q_{\rm inv}$ range. Within the scenarios considered, these results indicate that femtoscopic effects may need to be evaluated as a source of systematic uncertainty in future spin-correlation measurements. Their relative contribution in a selected $\Delta R$ interval may be estimated from the fraction of low-$q_{\rm inv}$ pairs in that interval,
the proper percentage of this contribution can be estimated by the fraction of low-$q_{\rm inv}$ range in the selected $\Delta{R}$ regions. 

{\bf Acknowledgments:}
This work is supported in part by the National Key Research and Development Program of China under Contract Nos. 2022YFA1604900 and 2024YFA1610700, and the Fundamental Research Funds for the Central Universities(XJ2026000302).
The numerical calculations were performed on a GPU cluster located at the Nuclear Science Computing Center, Central China Normal University (NSC3).

\bibliography{example} 

\end{document}